# A MOF-reinforced self-foaming sponge for mechanically robust triboelectric membranes with improved resistance to humidity

*Tianhuai Xu,*[1] *Pavel Kulyabin,*[2] *Fatih Uzun,*[1] *Alejandra Sophia Lozano-Pérez,*[2] *Ketan Pancholi,*[3] *Amit Kumar*[2] *and Jin-Chong Tan*[1,*]

[1]Multifunctional Materials & Composites (MMC) Laboratory, Department of Engineering Science, University of Oxford, Parks Road, Oxford OX1 3PJ, U.K.

[2]EaStCHEM, School of Chemistry, University of St Andrews, St Andrews KY16 9ST, U.K.

[3]The Sir Ian Wood Building, Robert Gordon University, Garthdee Rd, Garthdee, Aberdeen AB10 7GE, U.K.

[*]*Corresponding Author*: jin-chong.tan@eng.ox.ac.uk

**Abstract**

Porous triboelectric materials offer significant potential for enhancing the performance of triboelectric nanogenerators, yet their practical application is limited by structural instability and humidity-induced performance degradation. In this work, a bio-derived, sustainable polyamide containing disulfide linkages was developed to enable spontaneous formation of a porous dielectric without external templating. Hydrophilic MOF fillers comprising HKUST-1 crystals are incorporated within the porous matrix to reinforce the membrane structure and regulate moisture effects. Mechanical characterization demonstrates that HKUST-1 suppresses pore collapse and improves structural robustness under repeated deformation, while analysis of stress-strain behaviour reveals the critical role of pore stability in achieving stable triboelectric output. In addition, HKUST-1 mitigates humidity-induced charge dissipation by confining water molecules within its framework, giving enhanced triboelectric output stability and reduced performance degradation under increasing relative humidity compared to the neat porous system. This work demonstrates a strategy that integrates self-foamed porous structures, sustainable polymer design, and functional filler reinforcement to engineer mechanically robust and environmentally stable triboelectric systems.

**Keywords**: triboelectric nanogenerators, porous structures, metal-organic frameworks, composite materials, mechanical integrity, humidity stability

## Introduction

The growing demand for sustainable and self-powered electronic systems has stimulated extensive interest in triboelectric nanogenerators (TENG) as versatile platforms for mechanical energy harvesting and sensing. TENG devices operate based on the coupled effects of contact electrification and electrostatic induction, offering advantages such as structural simplicity, broad material selections, lightweight and efficient energy conversion at low frequencies [1]. Despite these merits, their practical deployment is constrained by performance instability under realistic operating conditions, particularly in humid environments and under repeated (high cycle) mechanical loading [2,3]. Moisture adsorption on triboelectric surfaces can induce charge screening and accelerate charge dissipation, while structural degradation under cyclic deformation compromises long-term reliability. These limitations underscore the need for material design strategies that will simultaneously ensure a long-term mechanical integrity and environmental robustness.

Porous dielectric structures have emerged as a promising approach to enhance TENG performance. The introduction of porosity reduces the effective modulus and hence mechanical stiffness of the material, enabling larger deformation under applied stress for enhanced contact compliance, particularly under low-force stimuli [4]. This mechanical adaptability facilitates improved charge generation compared to dense counterparts. The porous architecture also improves breathability and stretchability of the material, thereby enhancing comfort when deployed in wearable devices [5]. Moreover, the presence of internal voids offers a versatile platform for incorporating functional fillers, enabling an even higher output or additional functionalities within the triboelectric system [6]. For instance, Kou *et al.* incorporated fluorinated ethylene propylene (FEP) powder into a porous polydimethylsiloxane (PDMS) sponge and achieved 1.2 times higher output than the pristine porous film [7]. Vafaiee *et al.* fabricated a humidity sensor by infiltrating carbon nanotubes into a porous PDMS matrix, which resulted in a 52% enhancement in humidity sensing response with a linear sensing behaviour of up to 80% relative humidity (RH) [8].

Notably, porosity also introduces several competing limitations and material challenges. Common ways to introduce porosity include template-assisted approach [5,9,10], emulsion [11,12], phase separation [13,14], gas foaming [15], solid-state sintering [16] and 3D printing [17]. However, these foaming approaches are either time-consuming or limited to specific classes of polymers. For Nylon specifically, although known for its positive triboelectric properties, its intramolecular hydrogen bonding and semicrystalline structure result in a narrow processing window, which makes controlled pore formation extremely difficult [18,19]. In addition, the inclusion of air as a secondary phase lowers the effective dielectric constant, limiting charge storage capability [20]. Mechanically, porous architectures inherently suffer from reduced mechanical robustness, making them susceptible to pore collapse and structural densification under high stress, especially when smaller pore sizes are involved [21]. More critically, the interconnected pore network facilitates moisture ingress, leading to the formation of adsorbed water layers that promote charge leakage, which will significantly degrade the electrical performance of TENG devices under humid conditions [22].

To address the aforementioned challenges, incorporating functional fillers into porous matrices provides a viable pathway to simultaneously reinforce mechanical stability and regulate environmental effects. Metal-organic frameworks (MOFs), such as the hydrophilic HKUST-1, possess well-defined microporous structures and strong affinity toward water molecules [23]. While they are widely explored for humidity sensing, their ability to confine and localize water molecules within internal nanosized cavities presents an alternative function as moisture regulators in triboelectric systems. Water adsorption increases the dielectric constant of HKUST-1 crystals by nearly 20 times [24]. By temporary confinement of water molecules and enhancement of the dielectric properties, such hydrophilic fillers can mitigate charge screening and protect against humidity-induced performance degradation, while preserving the intrinsic mechanical compliance of porous structures [25].

Herein, we report a HKUST-1 integrated porous nylon-like membrane that achieves enhanced mechanical robustness and improved humidity tolerance for TENG applications. A bio-based, sustainable polyamide synthesized from renewable resources containing disulfide linkages was developed, which exhibits good thermal stability and can be chemically recycled under visible light photocatalysis [26]. The introduction of disulfide bonds also enables a self-foaming process to generate porous architectures without any external templating. This intrinsic foaming capability not only simplifies film fabrication but also offers a facile route towards environmentally benign porous materials. The resulting porous structure provides a framework for embedding HKUST-1 fillers, which act to reinforce structural integrity and stabilize electrical output under mechanical loading and humid conditions. We systematically investigate the mechanical behavior of both dense and porous membranes, examine deformation-induced structural evolution and its impact on triboelectric output, and establish correlations between mechanical properties and electrical performance through finite element modelling (FEM). Furthermore, the humidity-dependent triboelectric response was investigated to elucidate the role of HKUST-1 in suppressing moisture-induced charge dissipation.

This study establishes a direct correlation between mechanical properties and triboelectric performance in a porous system, and presents a composite material design strategy that addresses the inherent trade-off between mechanical compliance and environmental stability. The findings provide new insights into the development of mechanically resilient and environmentally robust porous TENG materials, with potential applications in wearable electronics and low-frequency mechanical energy harvesting.

## Methodology

*Materials*

Chemicals used for synthesis in this study are all commercially available. 4,4' Dithiodibutyrate and copper nitrate trihydrate were sourced from Sigma-Aldrich. Trimesic (benzene-1,3,5-tricarboxylic, BTC) acid was purchased from Thermo Fisher. Priamine 1074 was purchased from CRODA Netherlands BV.

*Synthesis of HKUST-1 MOF crystals*

HKUST-1 crystals were synthesized following a previously reported procedure with slight modifications [27]. Briefly, 0.875 g (3.6 mmol) of copper (II) nitrate trihydrate and 0.42 g (2 mmol) of trimesic acid were dissolved in 24 mL of a 1:1 (v/v) $H_2O$/ethanol solvent mixture. The resulting solution was transferred and sealed in a 50 mL Teflon-lined stainless-steel autoclave, and heated at 110 °C for 24 hours under solvothermal conditions. After cooling to room temperature, the blue crystalline product was collected and dried in vacuum at 120 °C to remove residual solvent.

*Synthesis of polyamide with disulfide bonds*

4,4' Dithiodibutyric acid (6.3 mmol) and Priamine 1074 (6.3 mmol) were weighed and added to a 50 mL flask equipped with a magnetic stirrer. The flask was refilled with argon, and the reaction temperature was gradually increased to 180 °C, followed by stirring at this temperature for 2 h. Next, the polymerization reaction was continued for 1 h under reduced pressure (1 mbar). The reaction was cooled to room temperature and then frozen in liquid nitrogen, and the polymer was mechanically broken down to remove it from the flask. Detailed synthesis procedures and structural characterisation can be found in reference [26].

*Preparation of the porous and nonporous films*

*N*-methyl-2-pyrrolidone (NMP) was used as the solvent to dissolve the polymer at 50 °C overnight. The solution was then degassed and brought back to 50 °C to make sure all polymer was fully dissolved and remained in the liquid phase. The polymer film was made through a spin-coating method. The prepared polymer solution was poured onto an indium tin oxide (ITO)-coated PET (polyethylene terephthalate) substrate and spun at 750 rpm for 15 s, followed by 1200 rpm for 15 s. The film was then slowly cured under ambient temperature and humidity level overnight to generate the porous membrane. The average thickness of the porous membranes was $49.0 \pm 1.8$ μm.

The dense film was prepared from the same procedure but cured at 50 °C instead. The average thickness of the dense membranes was $49.8 \pm 1.0$ μm.

*Infiltration of HKUST-1 crystals*

The as-synthesized HKUST-1 crystals were incorporated into the porous membrane *via* a vacuum-assisted infiltration process. Specifically, 20 mg of HKUST-1 crystals were dispersed in 1 mL of deionized water to form a uniform suspension. The pre-fabricated porous membrane was fully immersed in this suspension and subjected to vacuum. Upon gradual release of the vacuum, the differential pressure facilitated the infiltration of HKUST-1 crystals into the pores of the membrane. This process was repeated three times to maximize crystal loading within the porous structure. The MOF loading was estimated to be ~6 wt%.

*Fabrication of TENG devices*

The as-prepared membranes on ITO-coated PET substrates were used as the tribo-positive layers. On the negative side, Kapton tape was adhered to another piece of ITO-coated PET substrate. Copper wires were directly attached to both ITO electrodes and rubber sheets were placed underneath the PET substrates to improve conformal contact. The nominal contact area of the device was 2 × 2 $cm^2$.

*TENG measurement under contact-separation mode*

The TENG measurements were tested in a contact-separation mode employing a customised TENG test rig shown in Figure S8. The test rig consists of a vertical sample holder for sample attachment, a linear stage to control the gap distance from sample surface to counter electrode, and a load cell (Tedea Huntleigh Model 614) to record the impact force. The cyclic motion was controlled *via* a power supply, a function generator (GW Instek AFG-2105) and an electromagnetic shaker (LDS V201). In a standard oscillatory test, the TENG device was attached to the sample holder and then subjected to a cyclic loading of 2 Hz. The tapping amplitude was maintained at 2 mm and the maximum impact force was maintained at 60 N. The open-circuit voltage was measured on a digital oscilloscope (PicoScope 5444B) equipped with a 100 MΩ high voltage probe (Rigol RP1300H). The short-circuit current and transferred charge were recorded on a Keithley 6514 electrometer.

*Humidity stability test*

The humidity response of the TENG devices was evaluated within a sealed glove bag, as illustrated in Figure S13. A commercial hygrometer (RS PRO RS-172TK) was placed inside the bag to continuously monitor the relative humidity (RH). Prior to each measurement, dry nitrogen was purged into the glove bag until the humidity fell below 10% RH to establish a stable baseline. Subsequently, humid air generated from a humidifier was introduced into the chamber through an outlet tube at a constant flow rate to gradually increase and control the humidity level at a specific value of 10% to 80% RH. During the humidity modulation process, the output voltage and current of the device were recorded simultaneously to correlate electrical performance with a defined environmental humidity.

*Materials characterization*

The surface topography of the neat porous polyamide membrane was examined under a scanning electron microscope (Hitachi SEM). The morphology of the HKUST-1 crystals and the composite membrane was examined under a field-emission scanning electron microscope (FESEM LYRA3 GM TESCAN). Energy dispersive X-ray spectroscopy (EDS) was also performed on the same microscope to analyse the chemical composition. The X-ray diffraction (XRD) pattern of the crystals was obtained from a Rigaku MiniFlex diffractometer with a Cu $K_{\alpha}$ source (1.541 Å). Fourier transform infrared (FTIR) spectroscopy was performed on a Nicolet iS10 FTIR spectrometer equipped with an attenuated total reflectance (ATR) module. Raman spectroscopy was performed on a Renishaw InVia confocal Raman microscope equipped with a 785 nm laser, a 1200 l/mm grating and 5 mW laser power. The stress-strain curves were obtained from a KLA nanoindenter adopting a continuous stiffness measurement (CSM) method, where the stiffness of the sample is continuously recorded as a function of indentation depth. The detailed test procedures can be found in Figure S9 in Supplementary Information.

*COMSOL finite element modelling (FEM)*

The coupled mechanical and electrical behaviour of the porous TENG was modelled using the Solid Mechanics and Electrostatics modules in COMSOL Multiphysics (version 6.3). To reduce computational cost while retaining the essential deformation and charge transfer characteristics, a two-dimensional (2D) representative area element (RAE) was employed under quasi-static loading conditions. The pore geometry was defined based on nominal dimensions obtained from SEM micrographs, with the pore size selected such that the overall porosity of the model matched that of the real porous membrane. A 2D quarter-symmetry model was implemented with symmetric boundary conditions applied along the lateral and the bottom boundaries to represent the periodic nature of the porous structure and approximate the behaviour of the bulk material.

To investigate the influence of mechanical nonlinearity on electrical output, the surface charge density was assumed to remain constant throughout the simulation, while charge generation was allowed to occur in response to expanded contact area under compressive loading. For the pore closure analysis, contact interactions between opposing pore surfaces were implemented using the penalty contact formulation. Higher mesh density was imposed around the pore edges to better resolve the area in contact. (See Figure S11) A penalty factor ($P_\mathrm{n}$) of 1 was introduced to improve numerical stability and facilitate numerical convergence during rapid gap closure associated with pore collapse. When penetration is detected (the physical gap $g_\mathrm{n}$ becomes negative), a modified contact pressure ($T_\mathrm{n}$) is induced between the contacting nodes as follows:

$$T_\mathrm{n} = -P_\mathrm{n} g_\mathrm{n}$$

The simulation successfully captures the onset of conformal contact but stops before full closure due to the computational limits of localised mesh distortion at a higher strain.

## Results and discussion

### Foaming mechanism

As illustrated in the schematic shown in Figure 1a, the foaming process was achieved purely on a solvent-assisted approach, without the use of additional chemical blowing/templating agents. Figure 1b illustrates the proposed foaming mechanism. The film formation proceeds *via* a vapor-induced phase separation (VIPS) process. Upon exposure of the heated, concentrated polymer solution to ambient moisture, the hygroscopic NMP solvent absorbs water and initiates a solvent-non-solvent exchange, leading to phase separation into a polymer-rich phase and a solvent-rich phase. The polymer-rich phase undergoes solidification, facilitated by the annealing process of the disulfide bonds. Meanwhile the solvent-rich phase evaporates, creating voids within the polymer matrix and resulting in a porous structure [28]. The asymmetric exposure to ambient vapor during formation induces a pore size gradient across the membrane thickness, with larger pores predominantly formed near the surface, and progressively smaller ones towards the bottom due to slower vapor diffusion rate [29].

The pore size of the foamed films or membranes can thus be tuned by adjusting the polymer concentration, as shown in Figure S1. With increasing solution concentration, the pore size initially increases and then decreases. Although lower concentrations promote larger pore formation, excessively low concentrations lead to poor viscosity, resulting in non-uniform surface foaming. The decreasing pore size at higher viscosity can be attributed to the slower solvent–non-solvent exchange rate and higher bubble retention. 0.33 g $mL^{-1}$ solution was used for subsequent studies because it produces uniform foaming while retaining a relatively large pore size of 16.8 ± 4.2 μm, as shown in Figure S2.

Further parametric experimental studies revealed that both humidity and temperature play a key role in regulating the foaming process. As shown in Figure S3, foaming does not occur under low-humidity conditions due to an insufficient solvent-non-solvent exchange rate. Conversely, an excessively high humidity also hinders phase separation, as the solvent-rich phase evaporates less efficiently, leading to residual solvent entrapment within the pores, which makes the porous architecture less stable. Likewise, temperature also significantly affects the foaming process (Figure S4). At lower temperatures, the slow solvent-non-solvent exchange prevents the formation of a stable porous network, whereas at higher temperatures, enhanced polymer solubility results in a dense, non-porous film. We established that optimal foaming was achieved at approximately 20 °C and 40% RH, where a well-defined porous structure was obtained.

**Materials characterization**

The morphology of the as-synthesized HKUST-1 crystals was examined by SEM, as shown in Figure 2a. Most crystals are faceted and exhibiting a well-defined octahedral morphology with an average particle size of approximately 18.2 ± 4.4 μm. The crystallographic structure was further confirmed by X-ray diffraction. As presented in Figure 2b, the diffraction pattern displays characteristic Bragg reflections indexed to the (200), (220), (222), and (400) planes, which are in good agreement with the simulated pattern of HKUST-1. These results confirm the successful formation of HKUST-1 crystals with desired morphology and size.

Figure 2c shows the ATR-FTIR spectrum of the porous membrane, before and after infiltration of the HKUST-1 crystals (as a pore filler). The porous polyamide (PA) foam shows characteristic infrared (IR) absorbance bands around 1543 and 1637 $cm^{-1}$, corresponding to the amide II (combination of N-H bending and C-N stretching) and amide I (C=O stretching), respectively. In addition, the peak at 3305 $cm^{-1}$ is assigned to N-H stretching of the amide linker. These bands indicate good polymerisation and successful formation of the polyamide. As shown in Figure S6, for the porous and dense membrane, no significant spectral difference was observed, which indicates that there is no chemical reaction during the foaming process, and the chemical structure of the polymer remains intact in the porous state. Incorporation of HKUST-1 yields the composite termed HKUST-1@PA, where the intensity of the band at 1372 $cm^{-1}$ increases, which is assigned to the symmetric stretching vibration of the carboxylate group in the BTC linker. Other IR bands of HKUST-1 include peaks at 727, 1449 and 1649 $cm^{-1}$, which are assigned respectively to the aromatic C-H bending (727 $cm^{-1}$), and the asymmetric stretching of the coordinate carboxylate group (1449 and 1649 $cm^{-1}$) [30].

The FESEM image of the HKUST-1 infiltrated membrane is shown in Figure 2d. Due to overlapping size distributions between the crystals and the pores of the selected PA matrix, crystals smaller than the pore openings almost fully entered the pores, with some larger crystals partially exposed on the polymer surface. The energy dispersive X-ray spectroscopic (EDS) mapping in Figure 2e shows the element distribution within a region of HKUST-1 loaded pore. As shown in Figure 2f, other than the uniformly distributed elements of C and O across the chemical map, the presence of sulphur element in the polymer matrix confirms the presence of disulfide linker, while the copper atom confirms the successful infiltration of the HKUST-1 MOF crystals.

Figure 2g presents the Raman spectrum for the base PA polymer. The Raman peak at 1299 $cm^{-1}$ is assigned to amide III (C-N stretching and N-H bending). The characteristic Raman shifts at 509 $cm^{-1}$ and 631 $cm^{-1}$ are attributed to the S-S and C-S bonds, respectively, which confirms the presence of the disulfide group within PA. Other pronounced peaks include the Raman shifts around 1437 $cm^{-1}$ and 2900 $cm^{-1}$, which correspond to the bending mode and stretching mode of the aliphatic chain in the polymer backbone, which are also commonly found in other polyamides, as shown in Figure S7. The Raman spectrum for HKUST-1 shows distinct peak centered around 825 $cm^{-1}$ and 1006 $cm^{-1}$, corresponding to the out-of-plane C-H bending mode and the symmetric stretching mode of C=C in the aromatic ring [31]. The heatmap shown in Figure 2h shows the Raman spectra mapped onto a local area where a single crystal of HKUST-1 is embedded within a pore in the PA foam structure. By mapping the obtained Raman spectra to the spectra for the base polymer and the MOF crystal, there is a sharp boundary identified between the cell wall and the crystal, indicating minimum chemical interaction between the polymer substrate and the crystal. This result is consistent with the route by which the MOF crystals were physically infiltrated into the porous PA matrix.

**Triboelectric output**

The TENG devices adopt a sandwich structure, in which the prepared HKUST-1@PA membranes serve as the tribo-positive layer, while Kapton tape functions as the tribo-negative layer due to its strong electron-withdrawing tendency. The proposed working principle of the TENG device is illustrated in Figure 3a. When the two triboelectric layers come into contact, opposite surface charges are generated *via* contact electrification. Upon separation, an electric potential difference is established between the back electrodes, driving electrons to flow from the top electrode to the bottom electrode through the external circuit. When the layers approach each other again, the potential difference drops, resulting in electron flow in the opposite direction.

Figure 3b-d presents the electrical output performance of the TENG devices measured under 20 °C and 50% RH. Compared with the dense film, the porous foam exhibits reduced output performance. The open-circuit voltage decreases from 225.6 ± 0.9 V to 124.5 ± 0.4 V, the short-circuit current density declines from 1.44 ± 0.11 μA $cm^{-2}$ to 0.58 ± 0.11 μA $cm^{-2}$, and the charge transfer density reduces from 3.46 ± 0.03 nC $cm^{-2}$ to 1.83 ± 0.03 nC $cm^{-2}$. The reduced output is attributed to the large pore size, which lowers the effective dielectric constant of the material and decreases the effective contact area during operation. The void space within the porous

framework also lowers the air breakdown voltage within the material, leading to easier charge leakage [32]. Notably, after incorporation of HKUST-1 fillers, the open-circuit voltage, current density, and charge density recover to 152.3 ± 0.3 V, 0.98 ± 0.11 μA $cm^{-2}$, and 2.27 ± 0.03 nC $cm^{-2}$, respectively. The HKUST-1 crystals act as dielectric inclusions within the pores, partially restoring the effective dielectric constant of the composite membrane. In addition, the exposed MOF particles increase the effective interfacial contact area and promote charge transfer at the triboelectric interface. The frequency-dependent output in Figure 3e shows that the output is relatively stable under different operating frequencies. Figure 3f demonstrates that the energy harvested from the TENG device can be used to charge capacitors, where the charging speed went down as the capacitance increased.

The output performance under varying external load resistances is shown in Figure 3g. The foam-based TENG achieves a maximum power density of 529 mW $m^{-2}$ at an optimal external resistance of 10 MΩ. The operational stability of the device is presented in Figure 3h. Notably, the output voltage gradually increases with repeated loading cycles. This enhancement can be attributed to a creep-induced thickness reduction of the foam under prolonged cyclic compression, which enhances the electric field strength, thereby improving its long-term output performance.

**Mechanical robustness**

The representative load-depth ($P$-$h$) curves obtained from nanoindentation for the dense, porous and HKUST-1 loaded porous sample are shown in Figure 4a, which reveal the local contact stiffness (i.e. instantaneous slope of $\Delta P/\Delta h$) of the membrane under different strain conditions. Compared against the $P$-$h$ curve for the dense film, which exhibits a consistent and steep increase of stiffness with increasing indentation depth, the porous membrane displays a nonlinear response consisting of an initial compliant regime followed by progressive stiffening. Such behaviour is consistent with the classical deformation model of cellular solids described by Gibson and Ashby [33]. In the initial regime, the response is dominated by elastic bending of the cell walls, resulting in an approximately linear increase in load. This is followed by a plateau region associated with elastic buckling of the cell walls, which gives rise to reduced stiffness. At higher indentation depths, a sharp increase in stiffness is observed as the pores collapse and the material densifies, indicating the transition to compression of the solid framework. Stiffening usually takes place when the indentation depth reaches the pore diameter, which clearly suggests pore collapse events. The presence of hysteresis between loading and unloading curves further confirms the viscoelastic nature of the porous structure [34]. Upon incorporation of HKUST-1 fillers, the load-depth response becomes more linear and exhibits reduced stiffening at higher indentation depths. By integrating the area enclosed by the loading and unloading curves, the plastic work of the HKUST@PA sample was determined to be 69 ± 3 nJ, compared to 78 ± 3 nJ for the porous PA sample, which indicates a more elastic recovery and less energy dissipation during the unloading step [35]. This suggests that the MOF fillers act as mechanical reinforcements within the porous framework, suppressing cell wall buckling and delaying pore collapse. As a result, the porous structure is stabilized, and the mechanical response transitions toward a more elastic and less dissipative behavior under compression.

The stress-dependent electrical output of the corresponding TENG devices is presented in Figure 4b. The stress is determined as the peak force detected from the load cell divided by the nominal contact area of the sample. The dense TENG exhibits a typical response in which the open-circuit voltage increases linearly at low stress and gradually reaches a plateau at higher stress levels. The porous TENG, however, shows a more complex behavior. In the low-stress regime, the voltage increases linearly with stress. At intermediate stress levels, the output begins to plateau before entering a second regime of gradual increase. Beyond a critical stress threshold, a sharp rise in voltage is observed, followed by eventual saturation. In contrast, the HKUST-1 reinforced porous TENG displays a response similar to the dense sample, with no abrupt increase in voltage observed up to 300 kPa.

The similarity between the mechanical response of the porous structure and its triboelectric output suggests a strong coupling between mechanical behavior and electrical performance. To further elucidate the relationship between mechanical deformation and electrical output, finite element (FE) simulations were performed using COMSOL. A 2D representative element containing a single pore was analyzed, as shown in Figure S10. Under compressive loading, the cell walls oriented perpendicular to the applied stress experience concentrated stress, leading to deformation of the pore into an elliptical shape with localized stress intensification. When this stress reaches the yield point, until the polymer can no longer sustain elasticity at a higher stress level, the cell walls will buckle or even collapse and leads to non-linearity in the stress-strain curve. To account for this mechanical variation, experimentally derived stress-strain data were incorporated into the model (detailed input parameters are shown in Table S1) to represent the effective mechanical behaviour of the structure. In the model, it is assumed that the surface charge density is kept constant at $2\times10^{-5}$ C $m^{-2}$ (estimated from experimentally determined charge density), and the material is allowed to expand freely in response to applied stress. The separation gap between the contact layers is set to be 2 mm. It is shown that the simulated output of the TENG device scales almost inversely proportional to the stiffness of the sample. As shown in Figure 4d, under low stress level, the output increases almost linearly with increasing stress, as the relatively low stiffness of the porous structure allows efficient deformation under small loads. After the material reaches its initial yield stress, the concurrent change in thickness and surface area leads to nonlinear change in the capacitance level, which gives nonlinear response in the potential output at the same time. However, it is worth noticing that the non-linear behavior alone, in particular the strain-hardening response, does not lead to the jump in voltage at higher stress levels, but rather a plateau behavior. This behavior arises from the reduced incremental deformation at higher stiffness, where increasingly larger stresses are required to induce further structural changes, thereby limiting the variation in capacitance. Therefore, for a porous substrate, we ascribe the voltage jump to pore closure event instead, particularly due to the exposure of the pore valleys under extreme compression. Figure 4e shows the deformation of an open surface pore under compression at different strain levels. As shown in Figure S12, beyond a strain of approximately 0.13, an abrupt increase in contact length is observed, indicating the onset of pore collapse. This additional contact interfaces induced by pore closure creates additional surface charge and larger potential difference between the back electrodes [36]. This effect is supported by electrical simulations on the model before and after pore collapse, as shown in Figure 4f, where a significantly higher potential is observed at the same displacement after pore closure.

Overall, the simulation results are in good correlation with the experimental data, which indicates that the abrupt increase in TENG output originates not from intrinsic material nonlinearity, but from geometric instability associated with pore collapse. In the linear elastic regime, the compliant porous membrane deforms readily under compression. As the contact area between the triboelectric layers increases almost proportionally with the applied stress, the electrical output also exhibits a linear response. As the material transitions into the nonlinear regime, progressive strain hardening increases the effective stiffness of the porous framework, requiring higher stresses to yield additional deformation. Consequently, the stress sensitivity gradually decreases, giving rise to a nearly plateau region observed in the voltage response. Upon reaching the densification regime, widespread pore collapse leads to a sharp increase in contact area, which results in the observed abrupt increase in voltage. In contrast, the HKUST-1 reinforced structure resists pore collapses due to the mechanical support provided by the fillers, thereby suppressing this instability and eliminating the sudden voltage jump.

**Humidity stability**

To characterize the hydrophobicity of the films, contact angle measurements were performed, which is shown in Figure 5a. Compared to dense film, the averaged water contact angle of the porous film increased from ~90° to ~115°, which means the porous structure renders the film more hydrophobic. And infiltration with HKUST-1 renders the porous membrane to become effectively more hydrophobic, with the water contact angle further rising to 119.7°. Intriguingly, the HKUST-1-loaded porous film appears to be the most hydrophobic film and thus leads to better resistance to increased humidity.

The humidity-dependent electrical response is presented in Figure 5b. At low relative humidity levels up to 20%, both the dense and porous films exhibit stable output. In contrast, the HKUST-1 loaded porous membrane shows a slight increase in open circuit voltage from 215 V to 230 V. As the RH increases beyond 30%, the voltage output of the dense and pristine porous samples drops rapidly, approaching nearly zero at elevated humidity levels. In comparison, the HKUST-1-loaded sample shows a milder decline rate and retains a measurable output of 21 V at 80% RH, indicating improved tolerance to moisture. The charge transfer density at different humidity levels suggests the same trend. An initial enhancement in output is followed by a gradual decline as the humidity increases.

The proposed mechanism behind enhanced humidity tolerance for the HKUST-1-loaded porous membrane is illustrated in Figure 5c. At low humidity, isolated water molecules coordinate with the $Cu^{2+}$ paddlewheel sites and are temporarily confined within the internal cavities. Such adsorption effectively reduces the air fraction inside the pores and increases the local dielectric constant. The presence of these dielectric inclusions enhances the overall effective permittivity of the composite membrane and promotes surface charge retention, leading to the observed voltage increase [37].

When the humidity exceeds 20% RH but below 40% RH, progressive water condensation within the confined pore space occurs due to capillary effects, leading to the formation of thin water layers along the exposed surface. These adsorbed water films increase ionic mobility and facilitate the formation of percolative conduction pathways, which accelerate charge dissipation and recombination. For both the dense and the porous films, such effect leads to a decline in electrical output. However, for the HKUST-1 loaded porous membrane, the high output is sustained, as continued water uptake by HKUST-1 partially offsets this effect by contributing to dielectric enhancement, facilitating the framework to maintain a relatively high output up to 40% RH, where the average open-circuit voltage diminished by only 4%.

At even higher humidity levels, once the internal cavities of HKUST-1 become saturated with water, the MOF no longer effectively contributes to charge retention. The conductive pore-filling effect then dominates, resulting in a pronounced decline in output performance. When exceeding 70% RH, the voltage of all three samples decreases to below 80% of their initial values, indicating substantial humidity-induced charge leakage. However, due to dielectric enhancement of the water-confined HKUST-1 fillers and the higher hydrophobicity of the porous membrane, the HKUST-1 loaded porous membrane maintains a relatively higher output compared to the porous and the dense films.

As depicted in Figure S14, the moisture-exposed HKUST-1 loaded membrane can be readily reactivated by placing it in a sealed environment with desiccants to remove absorbed water. Figure 5d compares the voltage output of the sample before and after reactivation following exposure to 60% RH. No significant change in output is observed after the drying process to induce water desorption, indicating that water uptake within the porous structure is temporary and does not induce permanent degradation in moisture. This reversible behavior demonstrates good reusability and stability of the HKUST-1 loaded porous system.

**Conclusions and Outlook**

In this work, a bio-based polyamide containing disulfide linkages was developed as a self-forming porous dielectric for triboelectric nanogenerators. The resulting membrane exhibits relatively large pore size while maintaining favourable triboelectric characteristics. By incorporating HKUST-1 within the porous matrix, the structural integrity of the system is effectively enhanced, mitigating pore collapse and improving mechanical robustness under repeated loading. The correlation between stress strain response and electrical output demonstrates that maintaining pore stability is essential for achieving consistent triboelectric performance. In addition to mechanical reinforcement, HKUST-1 serves as a moisture regulating component due to its strong affinity toward water molecules. The confinement of water within the MOF structure suppresses the formation of continuous conductive pathways within the porous network, thereby reducing charge dissipation under humid conditions. As a result, the composite membrane exhibits improved humidity tolerance and more stable electrical output compared to the pristine porous system, particularly at intermediate relative humidity levels.

Overall, this work provides new insights into the development of mechanically robust and environmentally stable TENG materials, and offers a scalable pathway toward practical self-powered devices. Further optimization can be achieved by tailoring the loading fraction of HKUST-1 to balance mechanical compliance and moisture regulation. In addition, tuning the pore size and filler dimensions presents an effective route to extend device performance under more extreme humidity conditions. The versatility of this porous platform also enables the integration of additional functional fillers, opening opportunities for multifunctional devices. For example, stress-responsive fillers could be introduced to modulate mechanical behavior and enhance force sensing capability, while temperature-responsive materials may enable simultaneous thermal sensing. These directions highlight the broader potential of engineered porous architectures in advancing next-generation self-powered systems.

**Declaration of Competing Interest**

The authors declare no competing interest.

**Acknowledgements**

This work was supported by the UKRI Engineering and Physical Sciences Research Council (EPSRC) Award (TEGMOF EP/Z534146/1).

**Data Availability**

Data will be made available on request.

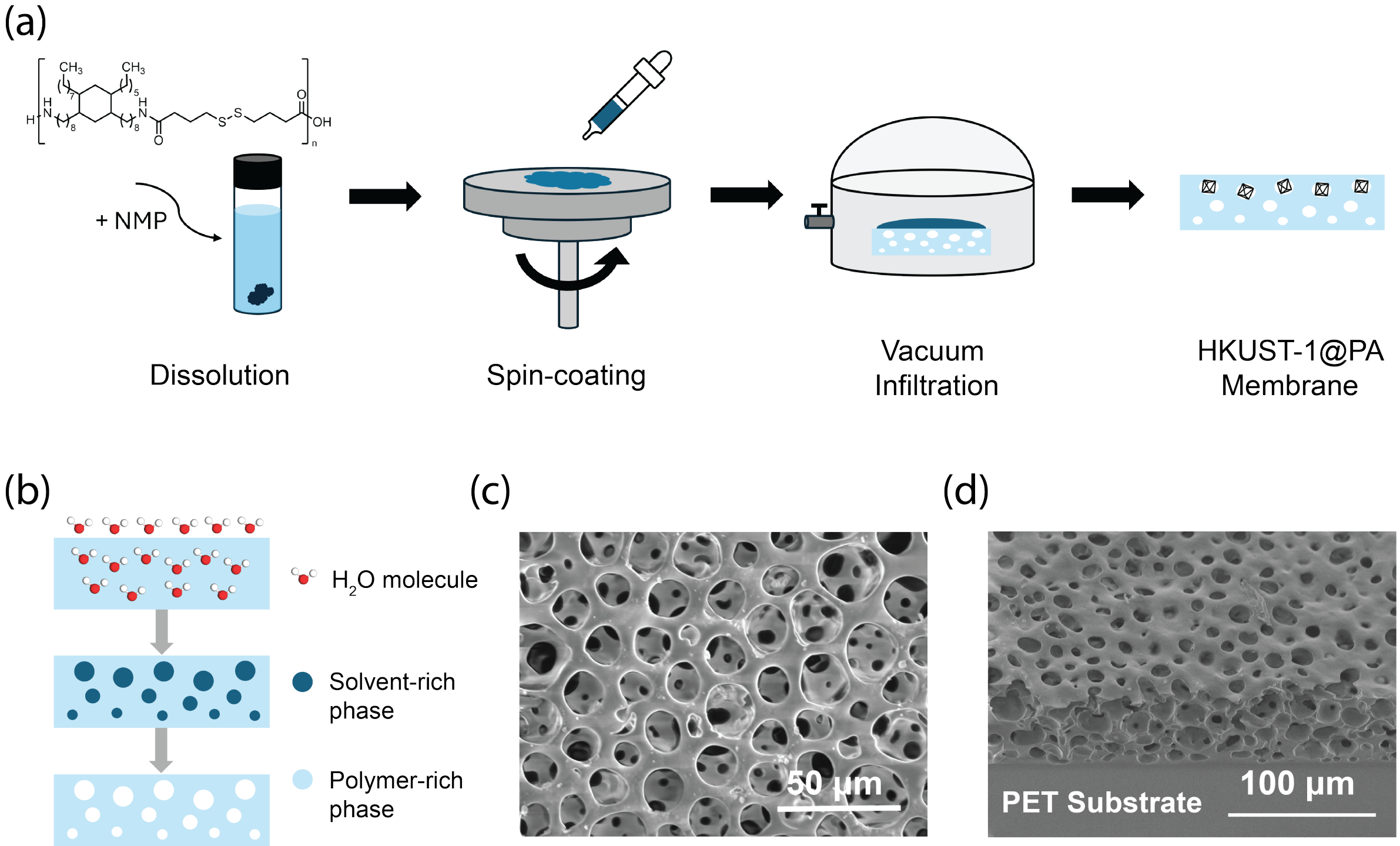

Figure 1. (a) Schematic of the foaming steps and the HKUST-1 impregnation process. (b) Illustration of the proposed foaming mechanism. (c-d) SEM micrographs of the top and the side views of the porous membrane cured from an NMP-dissolved polyamide solution with a polymer concentration of 0.33 g $mL^{-1}$.

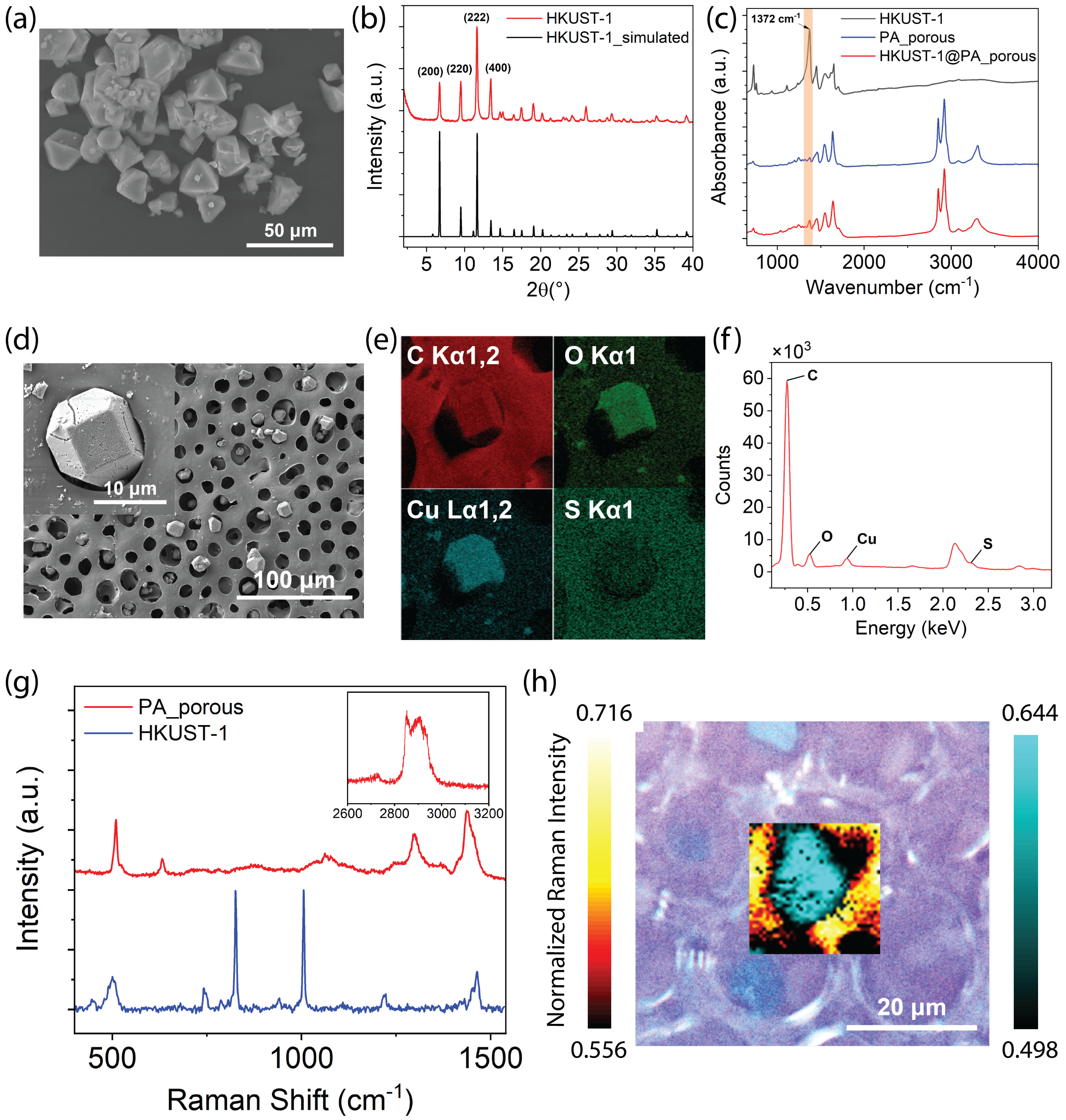

Figure 2. (a) SEM micrograph of the HKUST-1 crystals. (b) XRD patterns of the as-synthesized HKUST-1 crystals compared with the simulated pattern (CCDC code FIQCEN). (c) FTIR spectra of the HKUST-1 crystals, the neat porous membrane of polyamide (PA), and the HKUST-1 loaded porous membrane, designated as HKUST-1@PA. (d) FESEM image of the HKUST-1 loaded porous membrane. (e) EDS mapping of a region where a single HKUST-1 crystal is embedded within a pore. (f) Elemental analysis from the EDS mapping of panel (e). (g) Raman spectra of the neat polyamide material and the HKUST-1 crystals. (h) Raman mapping of a region with a single crystal surrounded by porous ligament, with the intensity of the corresponding component Raman spectra overlaid on top; the red scale corresponds to polyamide, and the blue scale corresponds to HKUST-1.

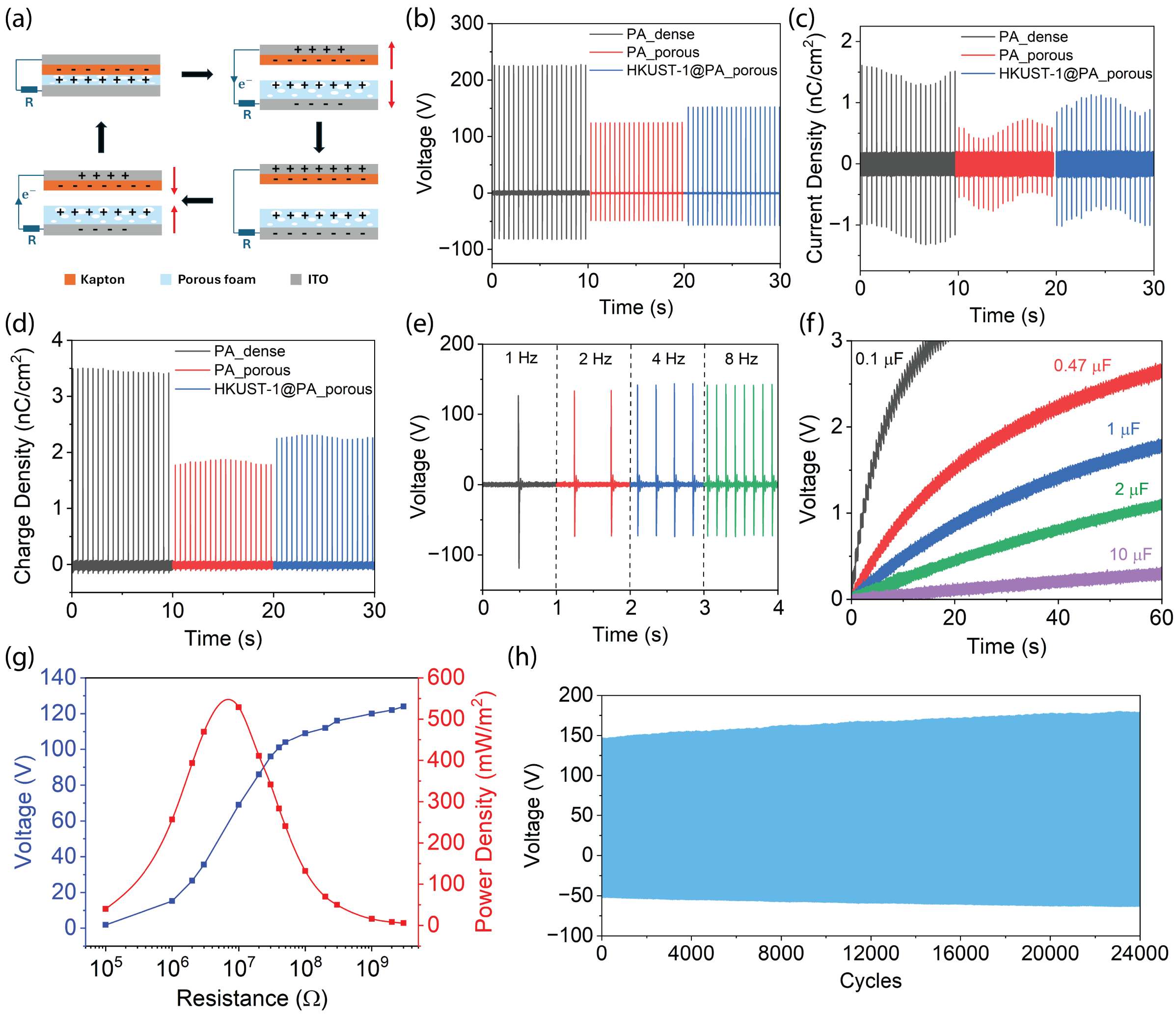

Figure 3. (a) Schematic of the proposed working mechanism of the contact-separation-mode porous TENG. (b-d) Comparison of the open-circuit voltage, closed-circuit current density and the charge transfer density of the dense, porous and HKUST-1-impreganated porous film. (e) The voltage output of the HKUST-1-loaded porous TENG at different operation frequencies. (f) The voltage profiles for charging different capacitors by the HKUST-1 loaded porous TENG operating at 2 Hz. (g) The peak voltage output and the corresponding power density of the HKUST-1 loaded porous TENG measured across a range of resistances. (h) Long-term stability of the HKUST-1 loaded porous TENG measured over 24000 contact-separation cycles.

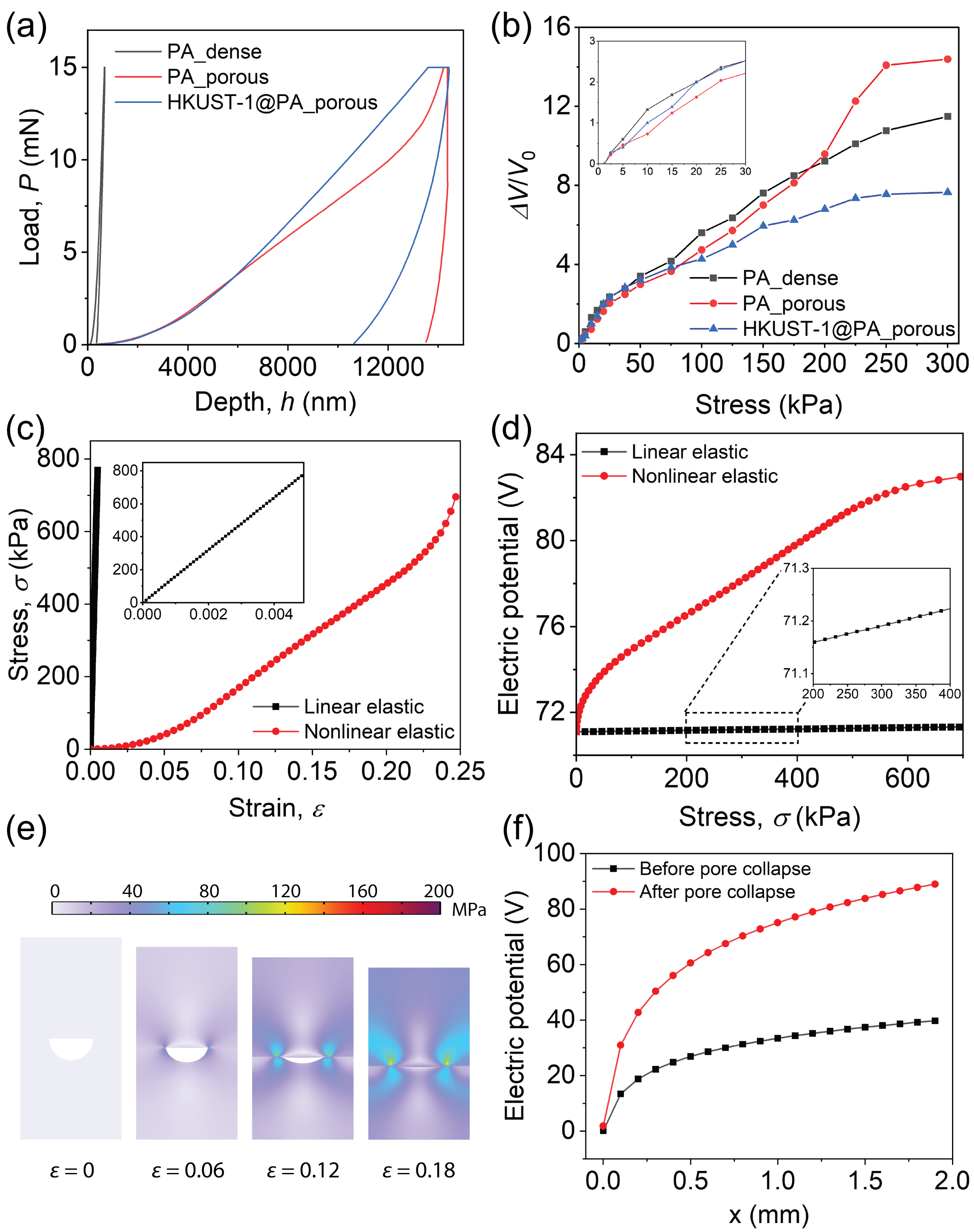

Figure 4. (a) The representative load-depth (*P-h*) curves of the dense, porous and HKUST-1 loaded porous membrane indented by a flat punch under an identical maximum load of 15 mN. (b) Relative change in voltage output of the dense, porous and HKUST-1 loaded porous TENGs under different compression stresses. (c) COMSOL finite element (FE) modelling of the stress-strain curves for dense (linear elastic) and porous (nonlinear elastic) samples. (d) FE-simulated electric potential under varying stress levels for both dense (linear elastic) and porous (nonlinear elastic) sample using the stress-strain relationships shown in panel (c). (e) Simulated

stress distribution maps surrounding a representative surface pore at different applied strain ($\varepsilon$) levels. (f) Simulated electric potential profiles at different gap distances between the top and bottom layers ($x$) before and after pore collapse.

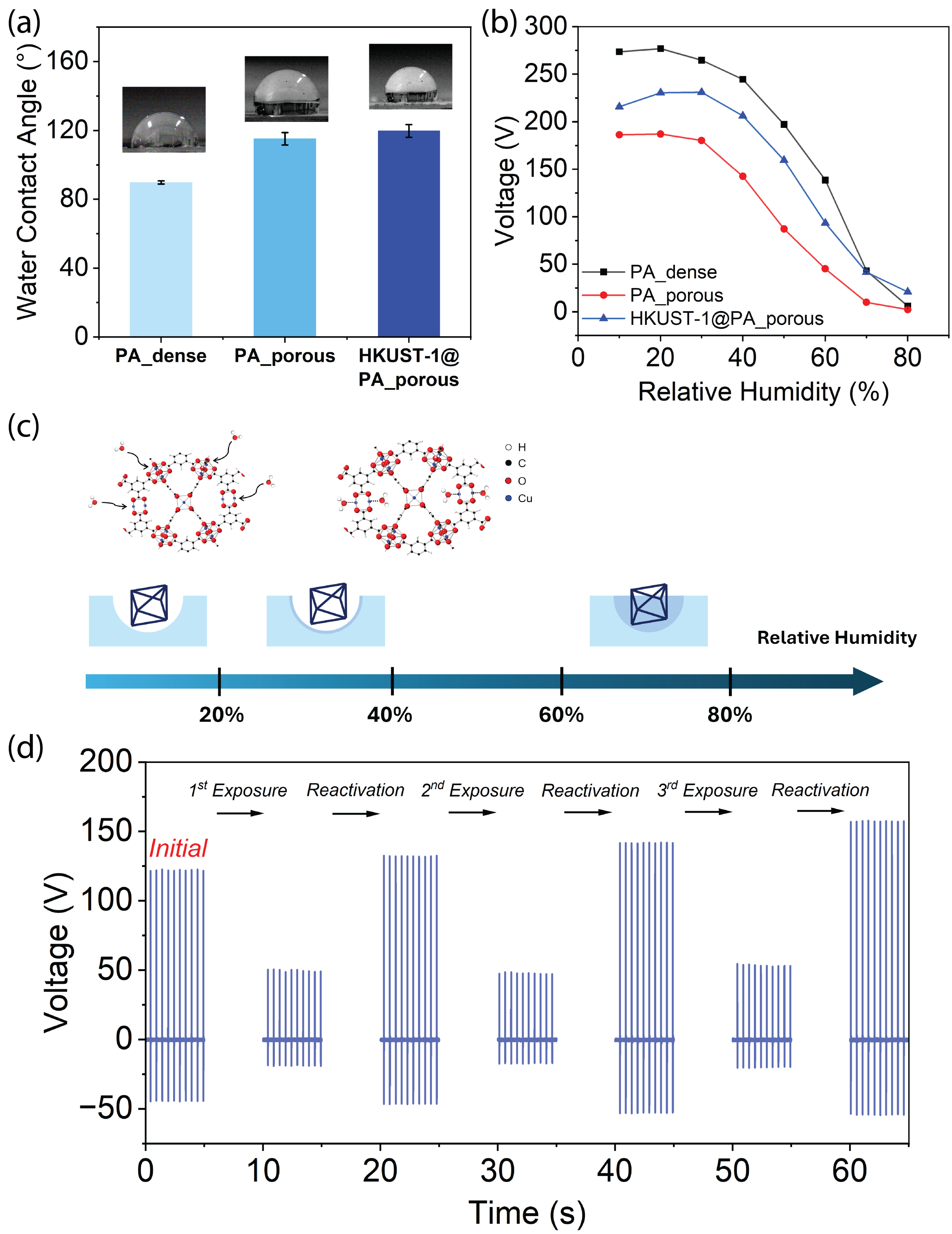

Figure 5. (a) Comparison of the water contact angles for the dense, porous and HKUST-1 loaded porous TENGs. (b) Comparison of the voltage profiles for the dense, porous and HKUST-1 loaded porous TENGs as determined under a rising relative humidity (RH) level. The error bars represent the standard deviation of the peak voltage measured over a 2 s interval and are small relative to the voltage differences observed between different RH levels. (c) Schematic of the water ingression mechanism at different RH levels; inset illustrates the adsorption of water molecules onto the open metal sites located on the copper paddlewheels of HKUST-1. (d) The voltage reversibility of the TENG device after 3 exposures to 60% RH followed by 3 reactivation steps.

# *Supplementary Information*

**for**

## A MOF-reinforced self-foaming sponge for mechanically robust triboelectric membranes with improved resistance to humidity

*Tianhuai Xu,[1] Pavel Kulyabin,[2] Fatih Uzun,[1] Alejandra Sophia Lozano-Pérez,[2] Ketan Pancholi,[3] Amit Kumar[2] and Jin-Chong Tan[1,*]*

[1]Multifunctional Materials & Composites (MMC) Laboratory, Department of Engineering Science, University of Oxford, Parks Road, Oxford OX1 3PJ, U.K.

[2]EaStCHEM, School of Chemistry, University of St Andrews, St Andrews KY16 9ST, U.K.

[3]The Sir Ian Wood Building, Robert Gordon University, Garthdee Rd, Garthdee, Aberdeen AB10 7GE, U.K.

[*]*Corresponding Author*: jin-chong.tan@eng.ox.ac.uk

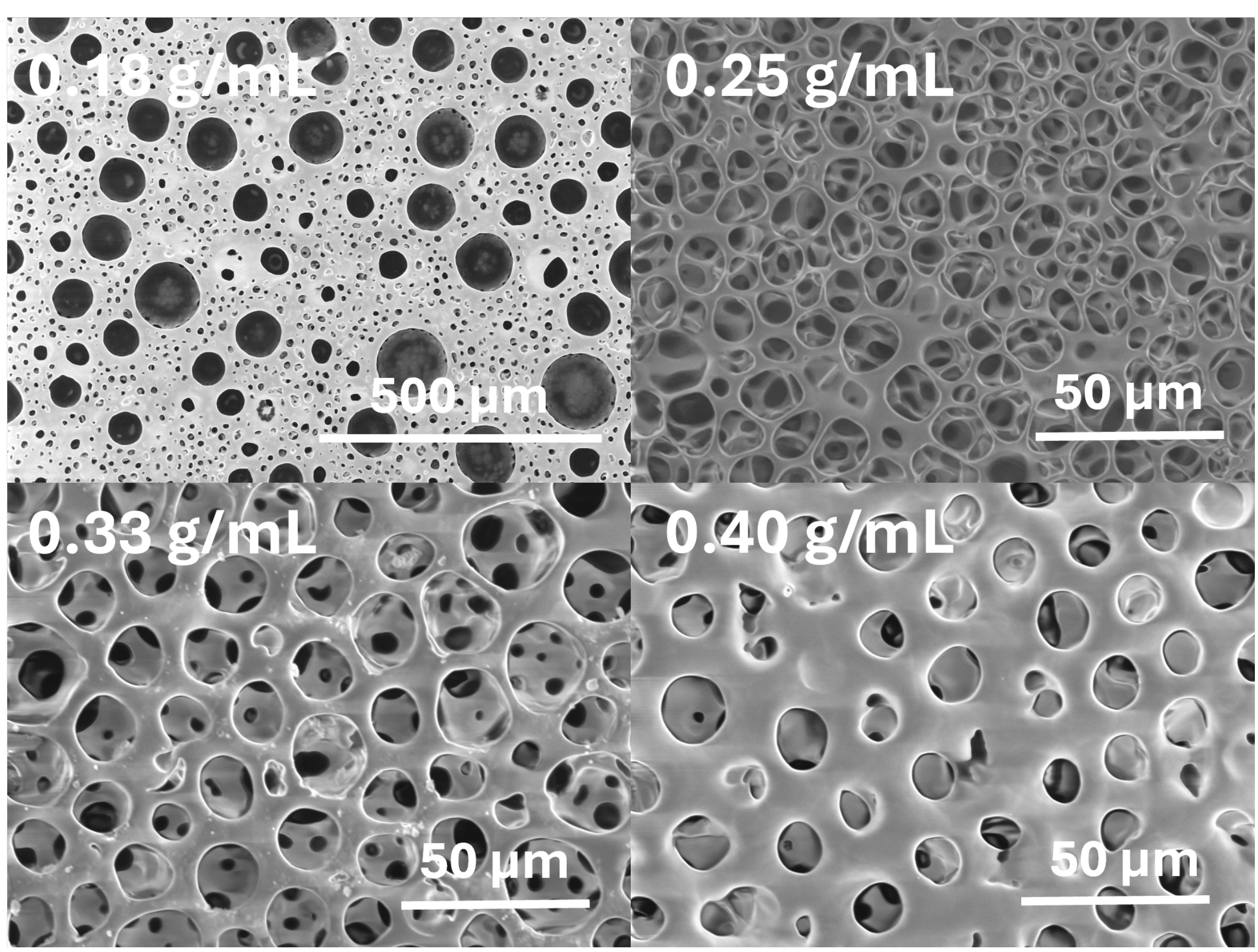

Figure S1. SEM micrographs of the foam samples made at different polyamide (PA) concentrations, with temperature at 20°C and relative humidity at 40%.

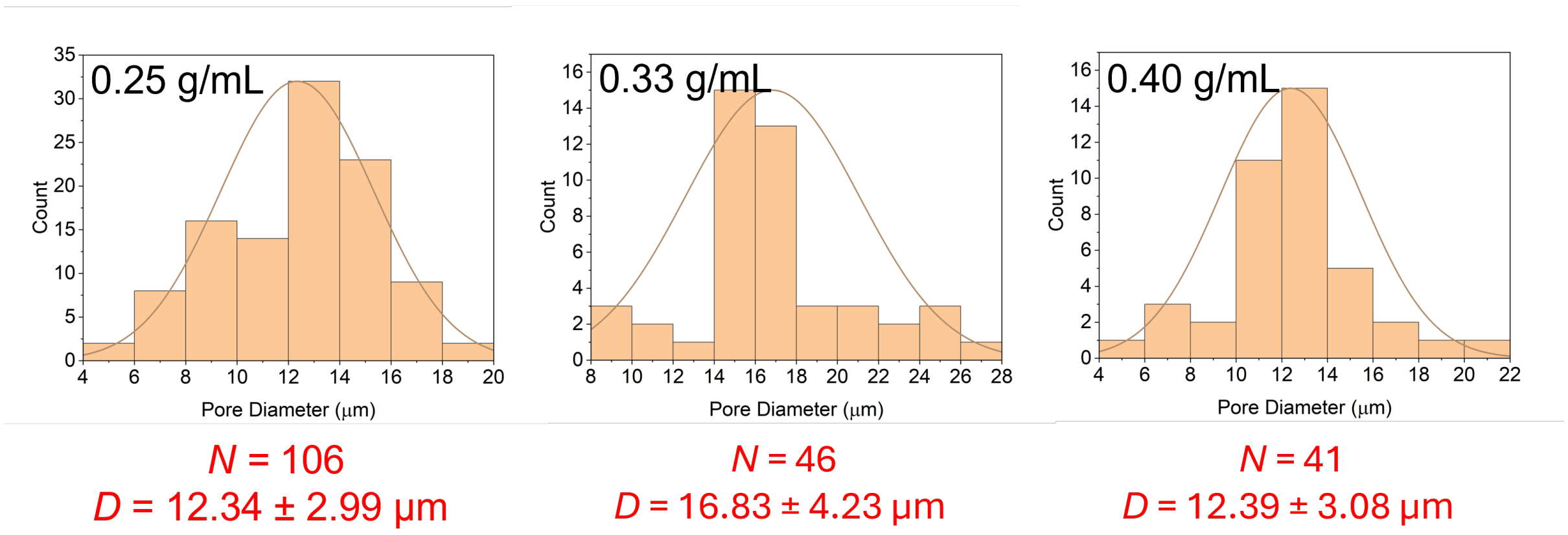

Figure S2. Pore size distribution of the foamed samples prepared with different polymer concentrations, with sample size (*N*) and average pore diameter (*D*) labelled at bottom.

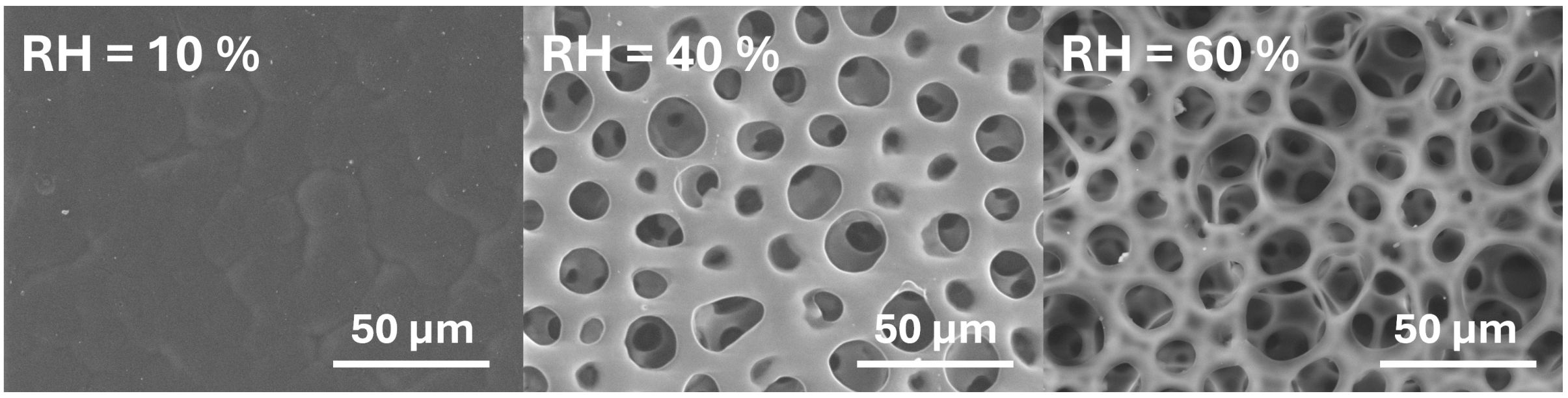

Figure S3. SEM micrographs of the foam samples (polymer concentration = 0.33 g/mL) made at different humidity levels, with the temperature maintained at 20 °C.

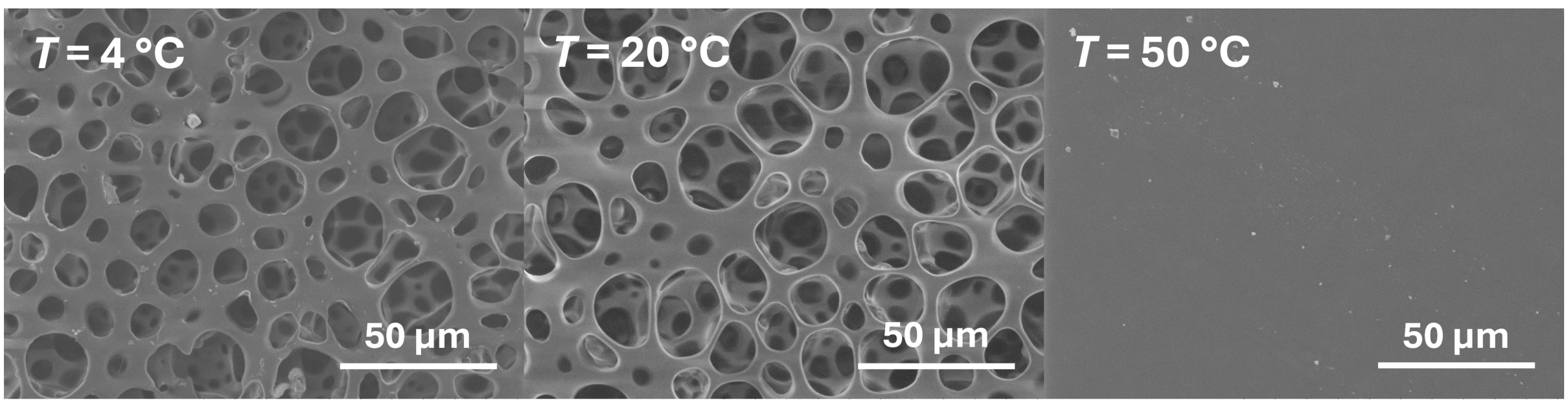

Figure S4. SEM micrographs of the foamed samples (polymer concentration = 0.33 g/mL) made at different temperatures, with the relative humidity maintained at 40%.

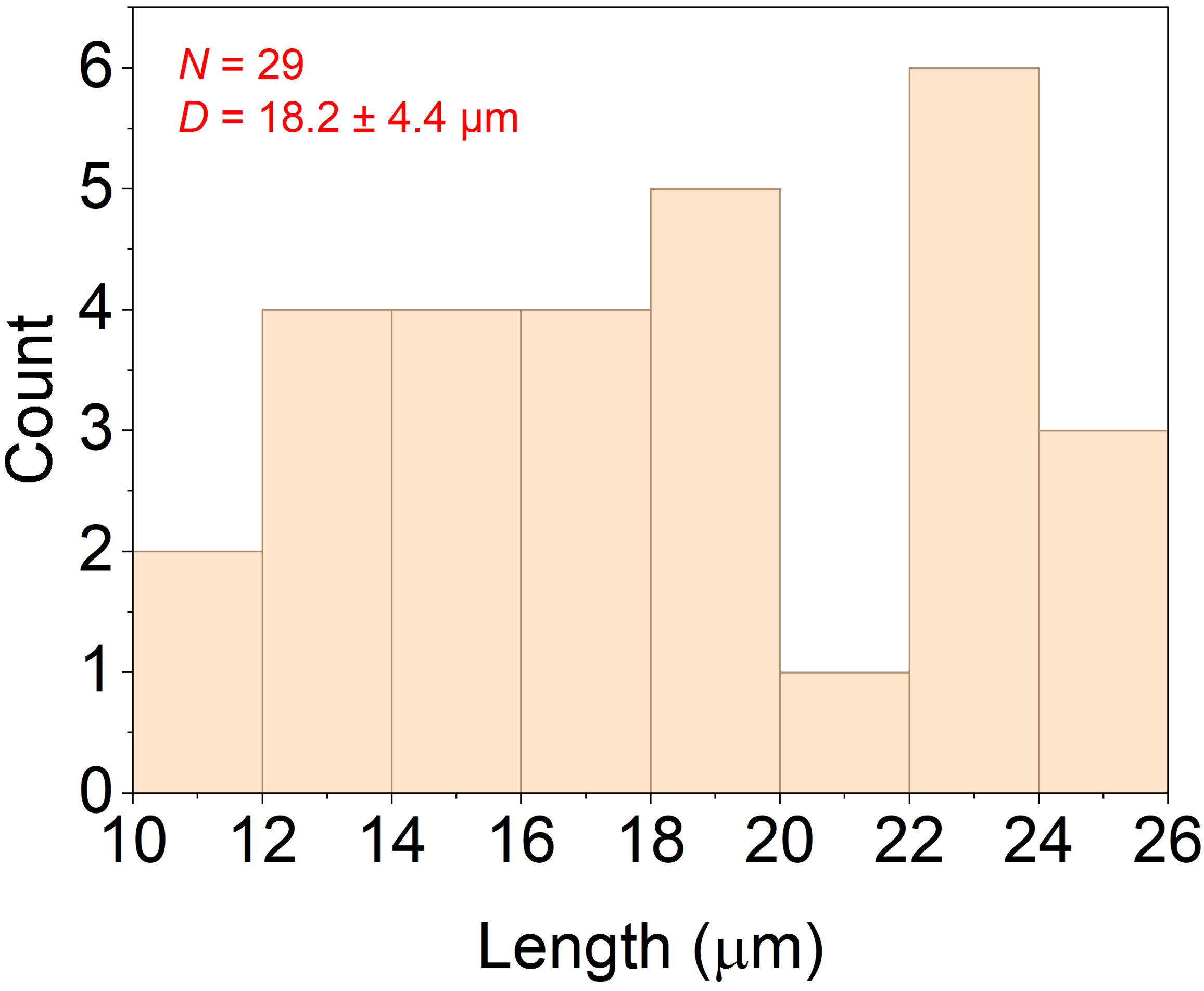

Figure S5. Particle size distribution of the as-synthesized HKUST-1 crystals, with sample size (*N*) and average particle size (*D*) labelled on top.

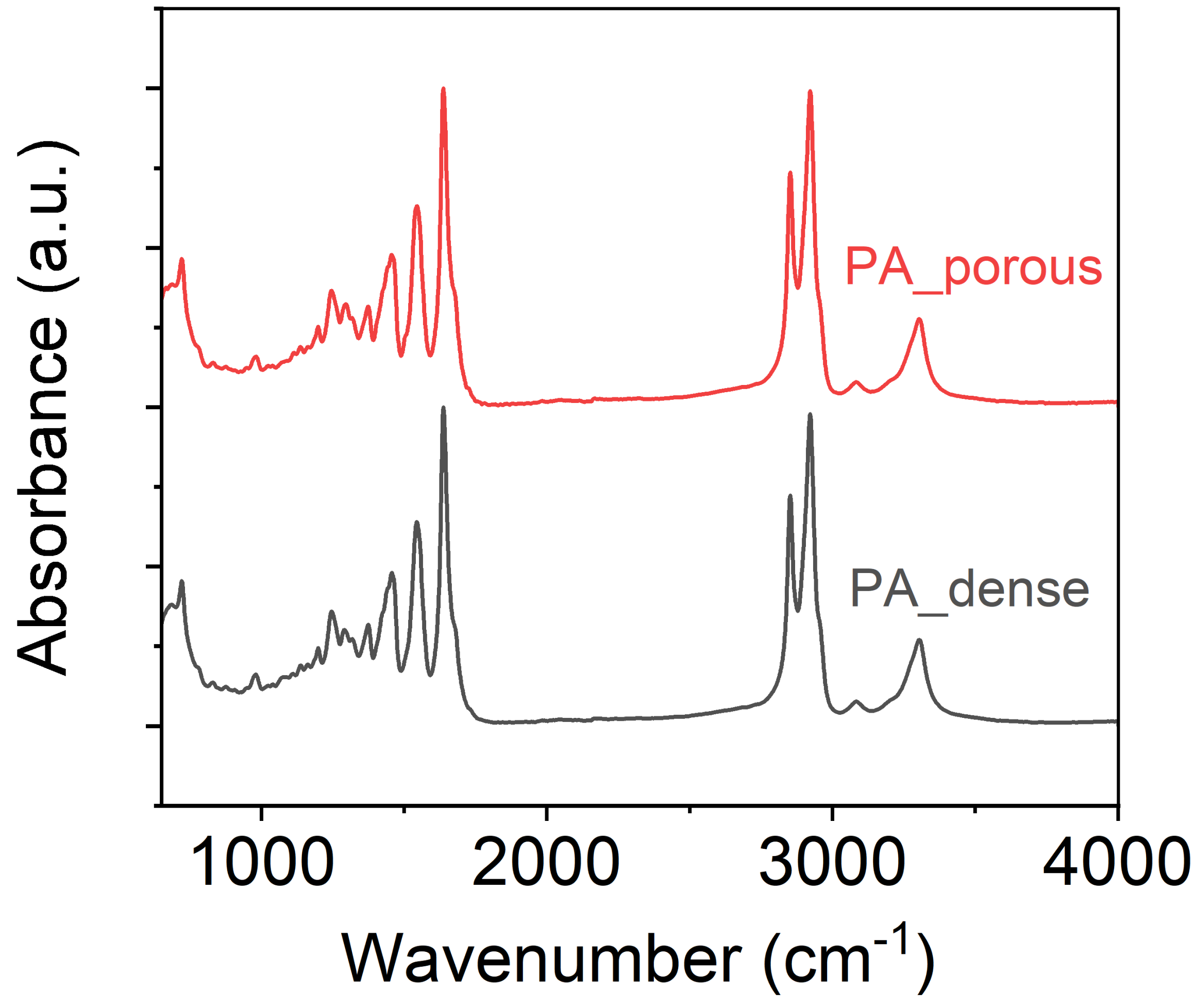

Figure S6. Comparison of the ATR-FTIR spectra of the dense and porous PA membranes.

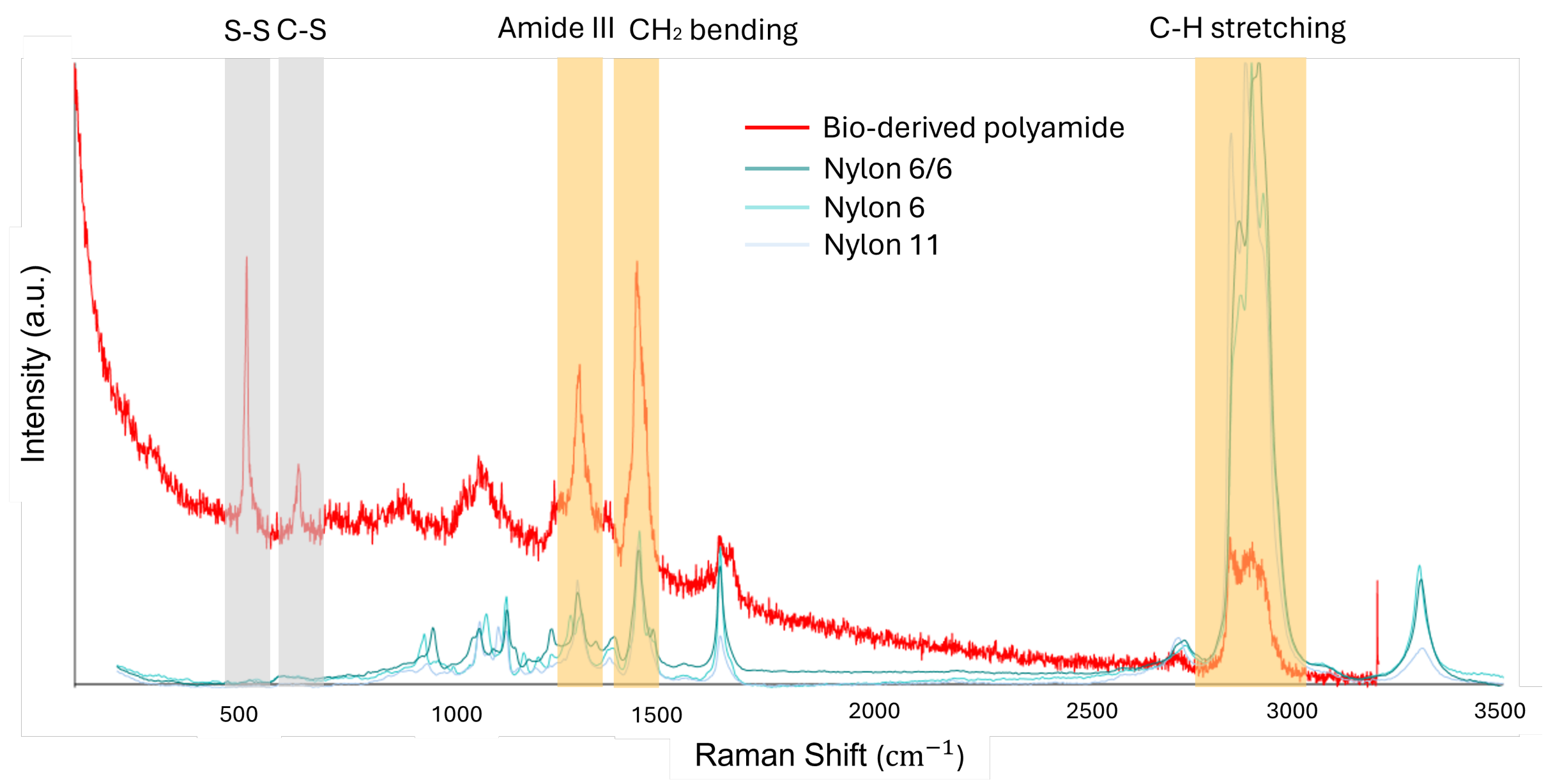

Figure S7. Comparison of the Raman spectra of the bio-based polyamide with other commercially available polyamides.

(a)

Load Cell
Magnetic Shaker
Sample Holder
Linear Stage

(b)

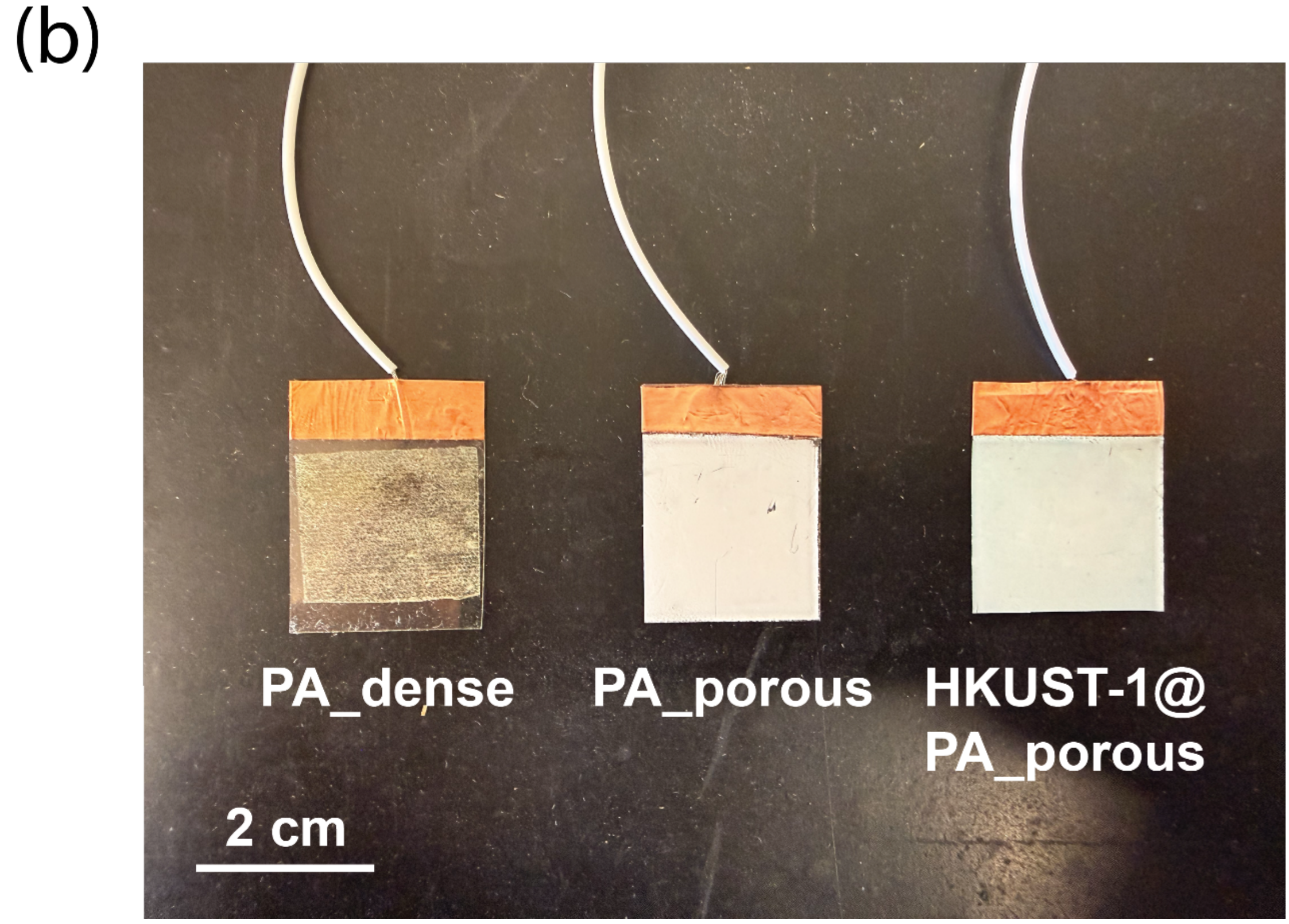

Figure S8. (a) Photo of the customised rig for TENG measurements under the contact-separation mode. (b) The prepared TENG devices using the dense, porous, and HKUST-1 loaded (HKUST-1@PA) porous samples.

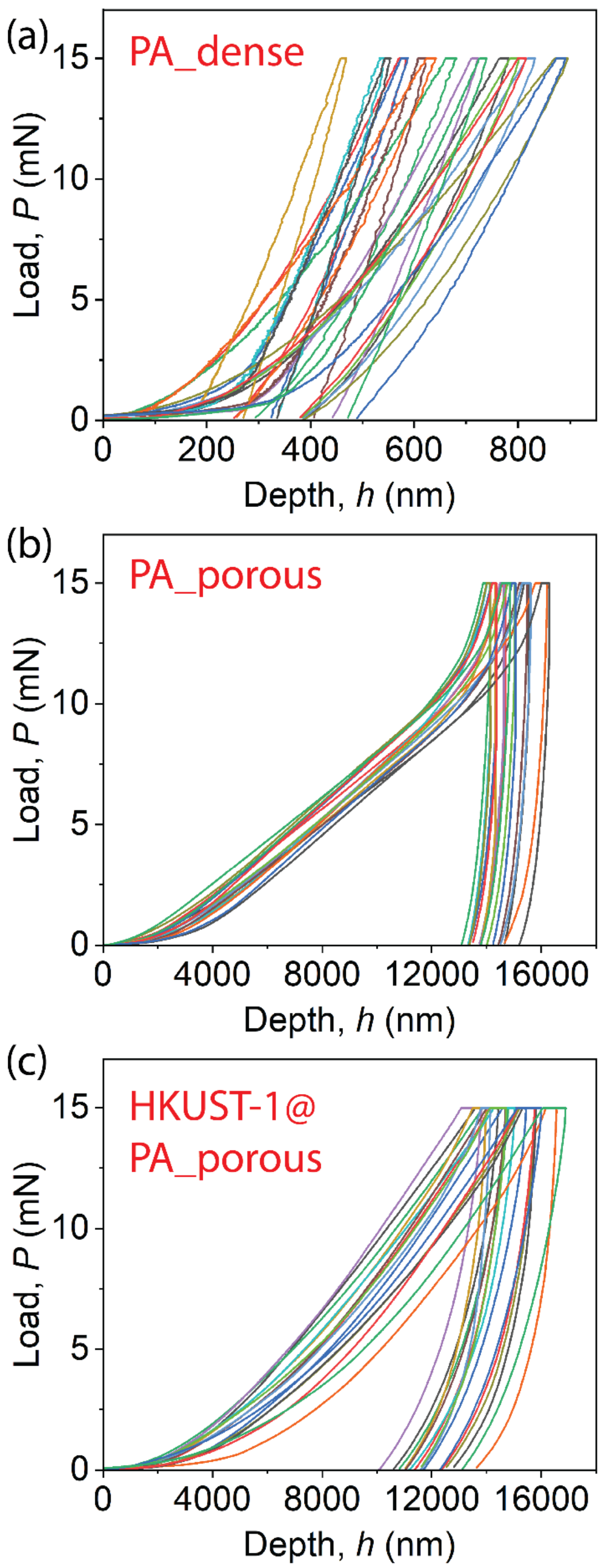

Figure S9. (a-c) Load-depth (*P*-*h*) curves of the dense, porous and HKUST-1 loaded porous samples, measured across 16 different locations. All the nanoindentation measurements were performed by adopting a continuous stiffness measurement (CSM) method. During the test, a flat punch with a diameter of 150 μm was used to cover as many pores as possible. The target load for the tests was set to be 15 mN. The holding time at maximum load was set to 1s. The tests were performed in a 2 × 2 array spaced by 300 μm in four distinct regions.

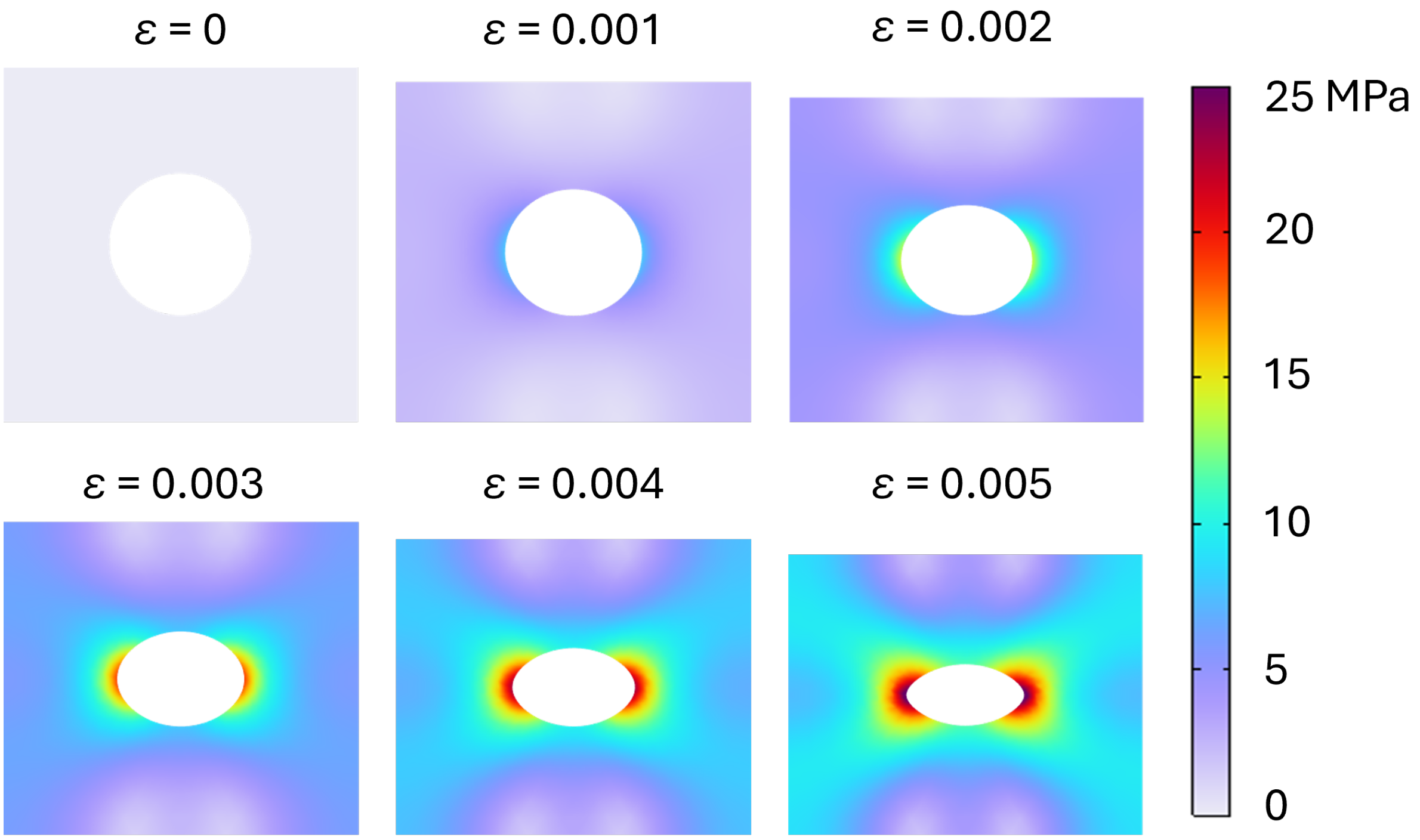

Figure S10. Stress distribution maps of the single-pore system under varying applied strain levels, $\varepsilon$. The initial diameter of the undeformed pore is 21.7 μm.

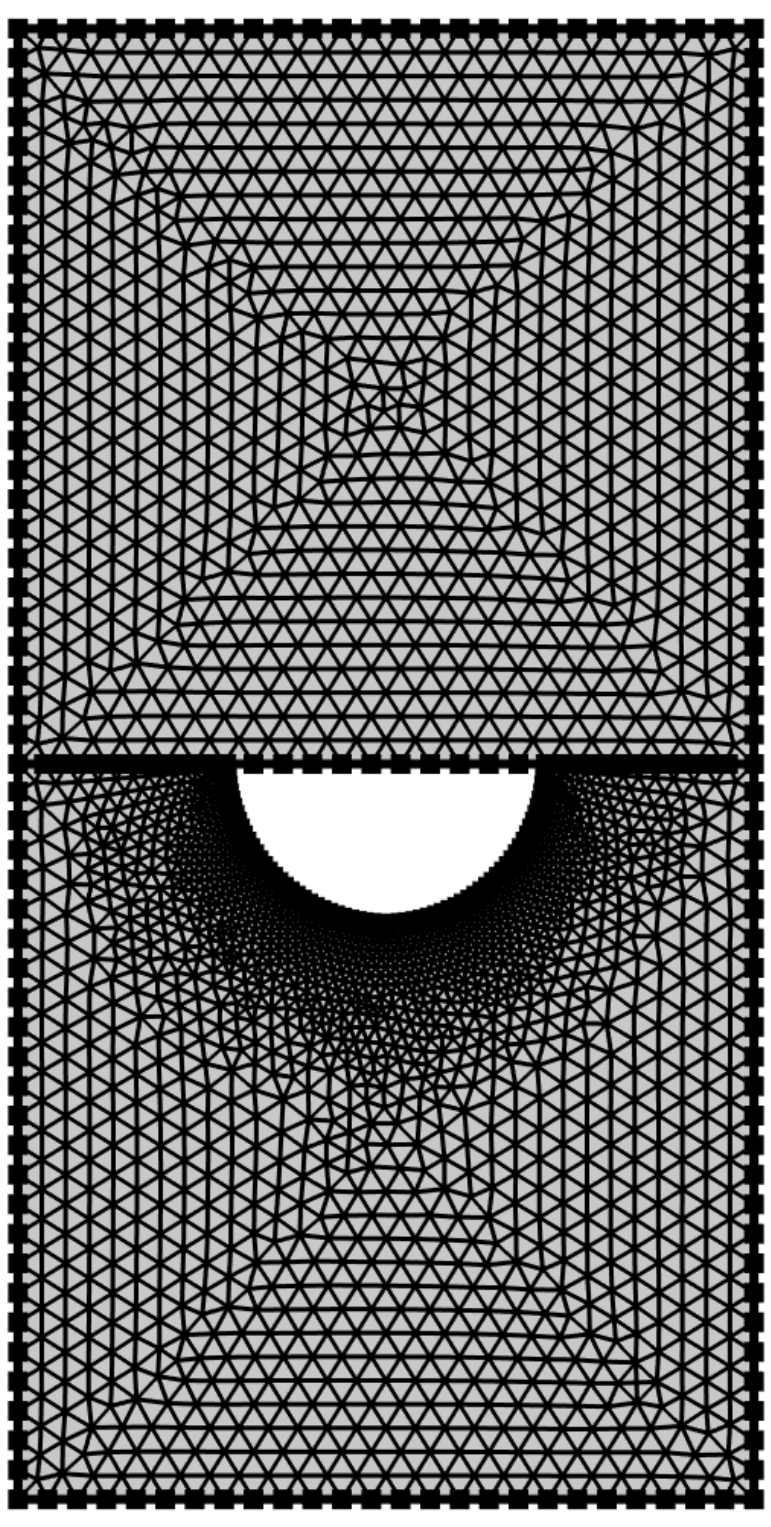

Figure S11. 2D FE mesh used for the pore collapse analysis in COMSOL. The pore diameter is set as 21.7 μm.

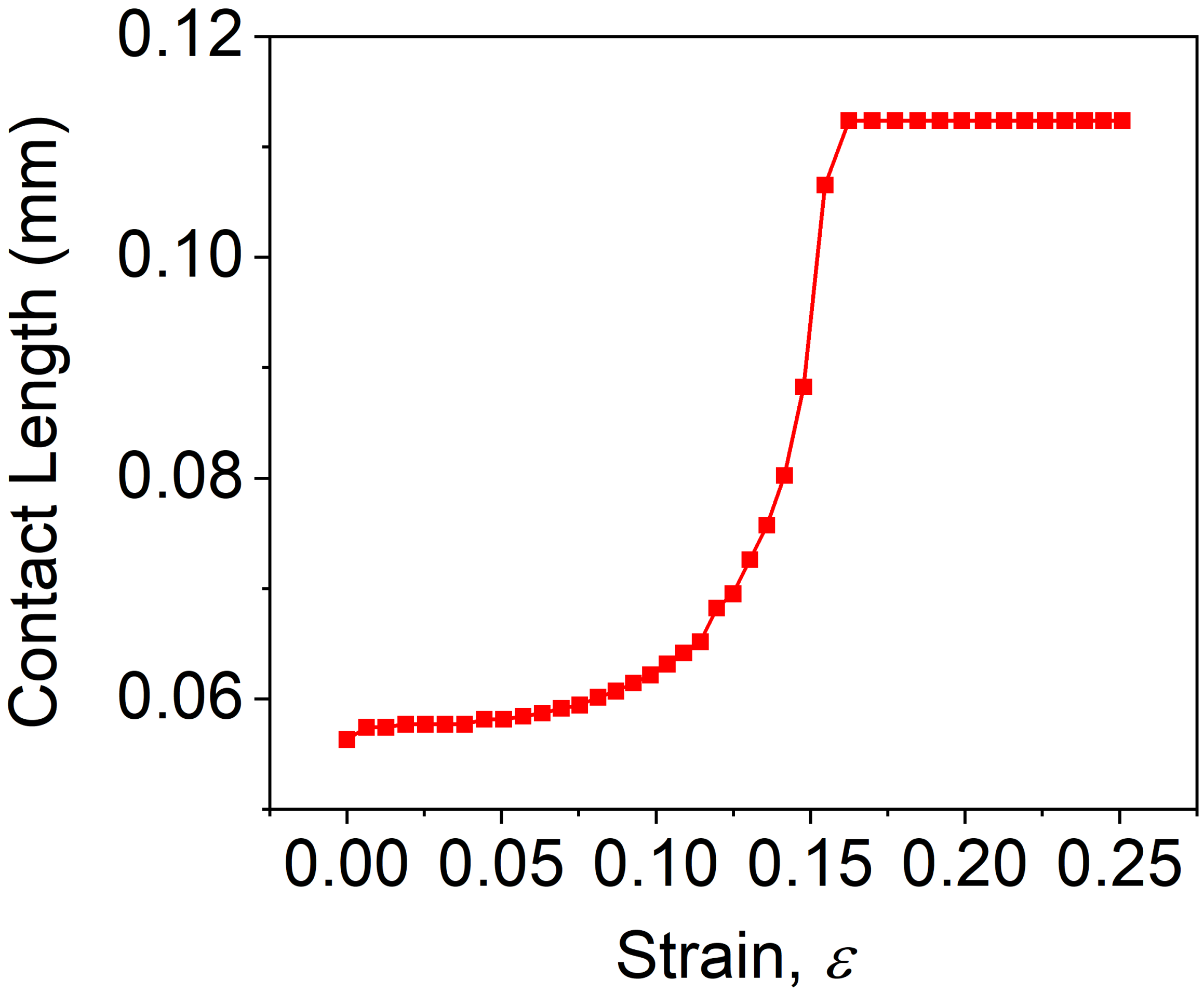

Figure S12. Evolution of the total length in contact between the top and bottom layers (contact length) at different strain levels from the pore collapsing event under FE simulation.

Table S1. The secant modulus data input into COMSOL to represent the measured stress-strain relationship of the porous sample. The secant modulus is defined as the average stiffness of the sample, and was calculated as the slope of the line from the origin to a specific point on the stress-strain curve.

| True Strain | Secant Modulus (MPa) |
|---|---|
| 0 | 0.00001 |
| 0.03145 | 0.46861 |
| 0.05278 | 0.86822 |
| 0.0682 | 1.19663 |
| 0.08046 | 1.47852 |
| 0.09115 | 1.72204 |
| 0.1015 | 1.92190 |
| 0.11146 | 2.09330 |
| 0.12106 | 2.24411 |
| 0.1305 | 2.37580 |
| 0.13985 | 2.49159 |
| 0.14903 | 2.59618 |
| 0.1581 | 2.69076 |
| 0.167 | 2.77822 |
| 0.17562 | 2.86151 |
| 0.18395 | 2.94212 |
| 0.19214 | 3.01818 |
| 0.20015 | 3.09091 |
| 0.20782 | 3.16364 |
| 0.21501 | 3.23874 |
| 0.22122 | 3.32433 |
| 0.22689 | 3.41387 |
| 0.23185 | 3.51017 |
| 0.23634 | 3.60967 |
| 0.2397 | 3.72364 |
| 0.24236 | 3.84555 |
| 0.24472 | 3.96995 |
| 0.2468 | 4.09659 |
| 0.24859 | 4.22608 |
| 0.25025 | 4.35596 |

Table S2. The detailed parameters of the mechanical simulation model

| | Material | RVE size | Young's modulus | Poisson's ratio | Density |
|---|---|---|---|---|---|
| Dense (Linear Elastic) | Polyamide | 50 μm × 50 μm | 150 MPa | 0.4 | 1150 kg $m^{-3}$ |
| Porous (Nonlinear Elastic) | Polyamide | 50 μm × 50 μm | User-defined (Interpolated secant modulus) | 0.4 | 1150 kg $m^{-3}$ |

Table S3. The detailed parameters of the electrical simulation model

| TENG layer | Material | RVE size | Pore diameter | Relative permittivity | Surface charge density |
|---|---|---|---|---|---|
| Tribo-positive | Polyamide | 50 μm × 50 μm | 21.7 μm | 4 | 2E-5 C/$m^{-2}$ |
| Tribo-negative | Kapton | 50 μm × 50 μm | N/A | 3.4 | -2E-5 C $m^{-2}$ |

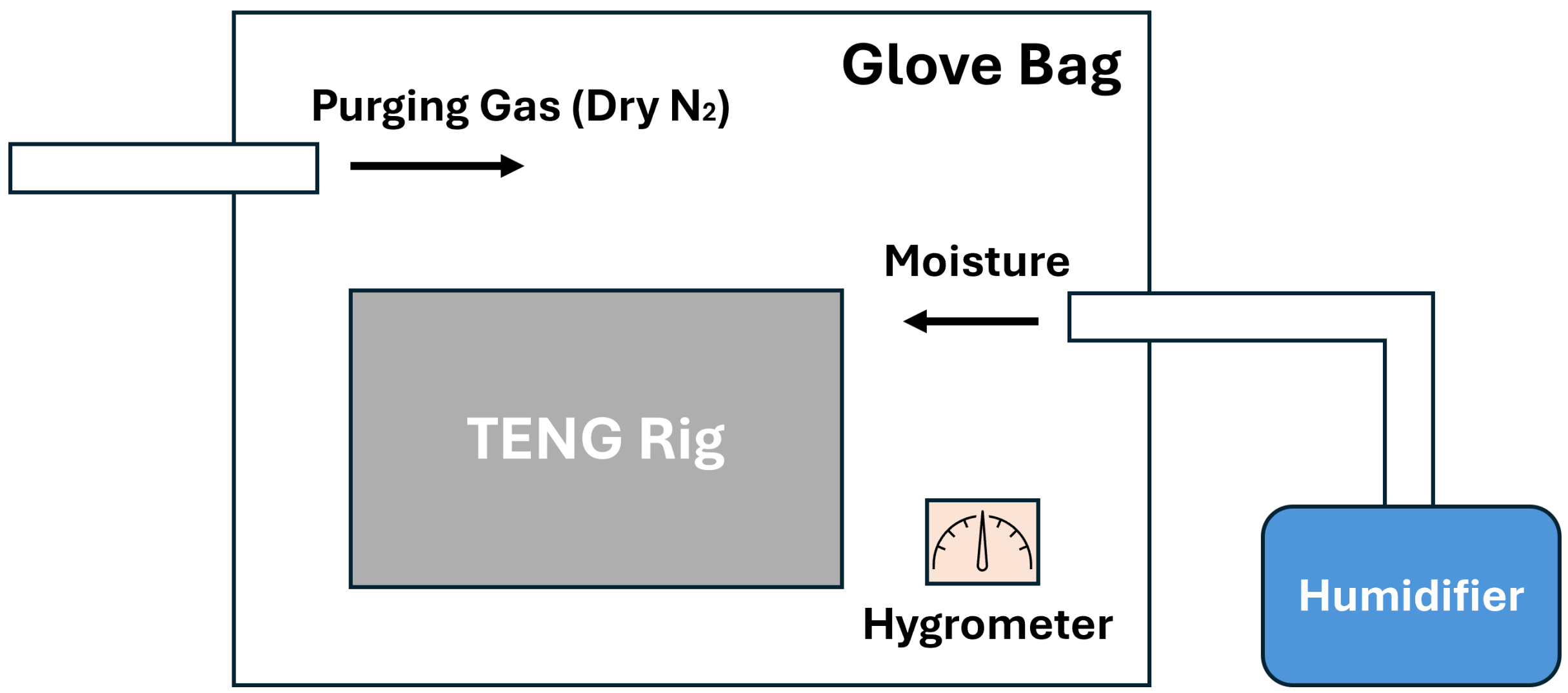

Figure S13. Schematic of the setup used for humidity control study.

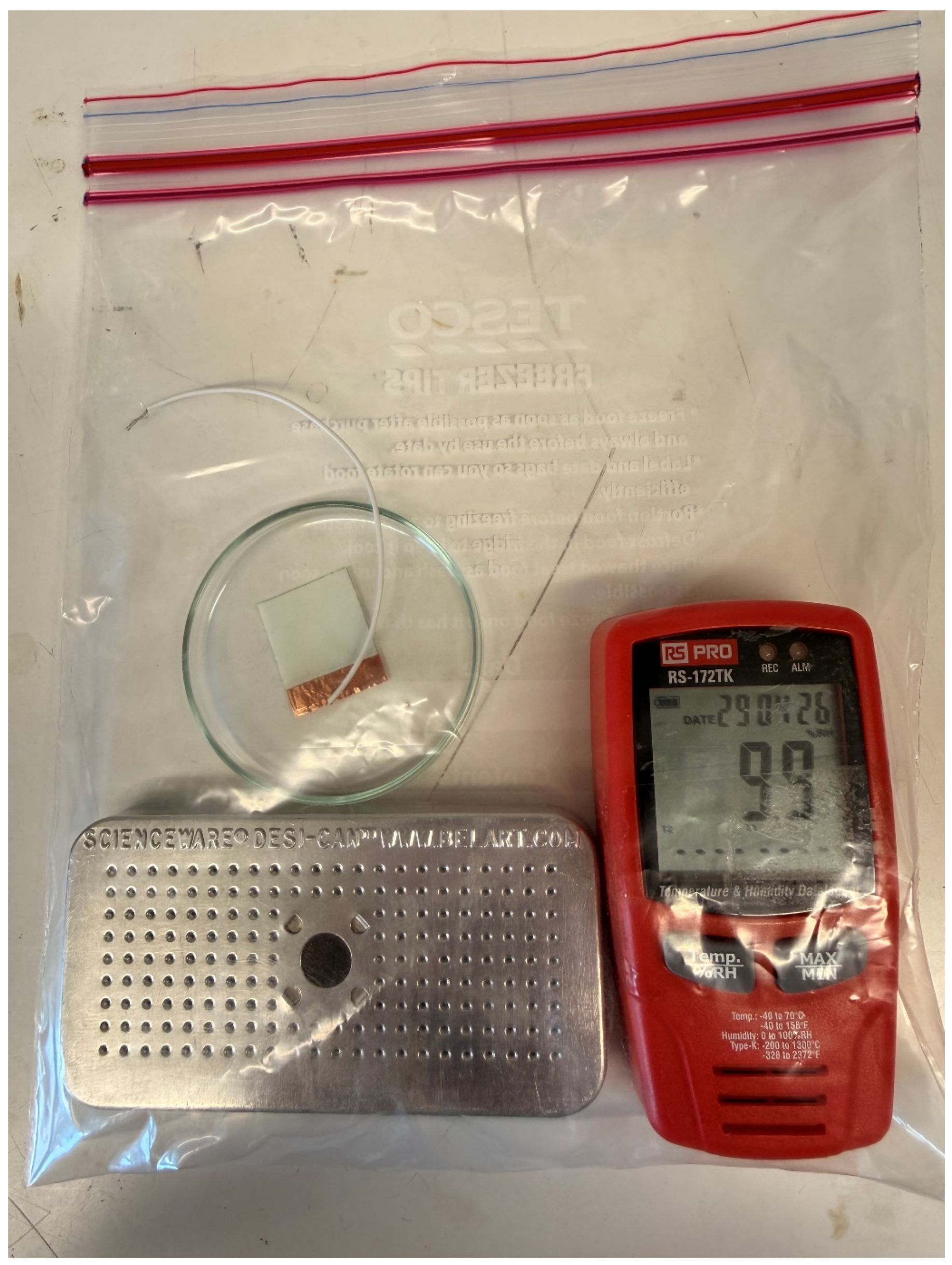

Figure S14. Reactivation of the HKUST-1@PA_porous sample inside the sealed bag with desiccants.